\documentstyle[preprint,aps,prb]{revtex}

\begin{document}

\draft

\title{
Hofstadter butterflies for flat bands
}

\author{
Hideo Aoki, Masato Ando$^*$ and Hajime Matsumura$^*$
}
\address{Department of Physics, University of Tokyo, Hongo,
Tokyo 113, Japan}

\date{\today}

\maketitle

\begin{abstract}

Hofstadter's diagram, or the energy spectrum against the magnetic 
field in tight-binding systems, 
is obtained for the models having flat (dispersionless) 
one-electron band(s) that have originally been proposed for itinerant 
spin ferromagnetism.  
Magnetic fields preserve those flat bands that 
arise from a topological reason, while dispersions emerge 
in a singular manner for the flat bands arising from interference, 
implying an anomalous orbital magnetism.  
\end{abstract}

\medskip


\narrowtext


Hofstadter's butterfly, or the Landau-quantized 
energy spectrum against the magnetic 
field in tight-binding systems, provides an intriguing 
example of fractal spectrum in the condensed matter physics.  
The quantum Hall effect for lattice fermions has also been 
discussed for the spectrum\cite{thouless}.  
Physically, the message is that when the magnetic field penetrating 
the unit cell of a two-dimensional (2D) 
lattice is $q/p$ in units of flux quantum, 
we have essentially a $p$-band system.  
Accordingly the scaling of the integer quantum Hall effect, 
for instance, exhibits a peculiar structure 
for $p\neq 1$\cite{anoma}.  

Although the situation might seem essentially the same 
for complex lattices with the unit cell containing several 
atoms, 
here we wish to point out that an interesting physics does exist 
when there exist {\it flat} (dispersionless) band(s).   
The flat band, or a macroscopic number of degenerate states, 
has appeared in the condensed matter physics 
from various contexts.  

First one concerns the spin magnetism in 
repulsively interacting itinerant 
electrons, as exemplified by the Hubbard model. 
It has become increasingly clear that only at, or possibly around, 
the singular limit of infinite interaction and infinitesimal 
doping from a half-filled band does a ferromagnetism 
appear.  
Lieb\cite{lieb} then pointed out that we can realize a ferrimagnetism, 
for arbitrary strength of the Hubbard $U$ at half-filling, 
if a bipartite lattice with nearest-neighbor 
transfers has different numbers, 
$n_{\rm a} \neq n_{\rm b}$, of a and b 
sublattice sites in a unit cell.  
In this situation 
$n_{\rm a} - n_{\rm b}$ flat band(s) appear, so that the ferromagnetism 
resides on flat bands. 

This is in accord with the 
`generalized Hund's coupling'\cite{kusakabe}, 
which dictates that electrons on the Fermi surface should be 
fully spin-polarized for any $U$ ---  a 
macroscopic number of states lying on 
the Fermi energy will then imply a bulk magnetization.   
Curiously, each `Wannier function' on the flat band cannot be confined to a unit cell contrary to a naive expectation 
that a dispersionless band should come from disjointed states.  
We can in fact identify the overlap of the Wannier states as 
an intuitive reason why the spins are ferromagnetically 
coupled\cite{kusakabe,mietas}.  
This reminds us of the fractional quantum Hall system, where 
the quantum-liquid ground state is fully spin-polarized 
due to the exchange interaction among the orbitals 
in a Landau level, a peculiar `flat band' arising itself from magnetic 
fields.  There, orthogonalized `Wannier states' cannot be constructed 
either\cite{thouWann}.

The model is extended by Mielke\cite{mielke} and by Tasaki\cite{tasaki}, which introduce 
distant-neighbor transfers to prepare flat band(s), on which 
spins align.  
Since the flat band is a result of an interference 
among the nearest-neighbor and more distant 
transfers, we may call this class the flat band due to interference.  
By contrast, Lieb's class may be called the flat band due to topology, 
since only the manner in which the sublattices are interlocked 
matters.  

A class of flat bands has also been conceived in the context of 
`lateral superstructures' 
that have superperiods of atomic dimensions along 2D 
directions. 
These are envisaged to be realized in 
organic $\pi$-electron materials such as the `long-period graphite' 
(with period $\sim$ of a few tens of \AA ), once alleged to be obtained 
in an attempt to fabricate fullerene\cite{chapman}.  
We can use the group theory\cite{shimaPRL} to classify 
all the atomic configurations with a superperiod into semiconducting, semimetallic and metallic  classes. 
The superperiod such as super-honeycomb structures 
{\it enforces}, in some classes, the existence of 
flat bands on 
top of dispersive ones, which is a systematic 
realization of Lieb's case.  

Now, a natural question is what will happen to the 
flat bands when a uniform magnetic field is applied.  
We can in fact expect intriguing phenomena, such as 
the orbital magnetism as 
in the `ring-current effect' 
in fullerene\cite{elser}.    
In the present paper we show that the Hofstadter butterfly 
for the flat-band systems 
reveals that the magnetic field makes the flat bands 
remain flat, sandwiched between 
usual Hofstadter butterflies, for the flat bands arising from topology, 
while the flatness is lifted in a singular manner for the flat bands 
arising from interference.  
These imply that not only the spin magnetism but 
the orbital magnetism are intriguing in flat-band systems.

For the tight-binding model on complex lattices 
we consider for convenience 
a rectangular unit cell of $L_x \times L_y$ 
(which is twice the original unit cell 
in the case of honeycomb systems).  
The strength of 
the magnetic field, $B$, applied perpendicular to the system 
is characterized by 
$\tilde{B} \equiv BL_xL_y/\Phi_0 = q/p,$ 
where $\Phi_0=h/e$ is the flux quantum and 
the field is called rational when $p,q$ are 
integers.  

The magnetic field is incorporated in the transfer energy, $t_{ij}$, 
in a usual manner through Peierls's phase as 
\begin{equation}
  t_{ij} \rightarrow e^{i\phi_{ij}} t_{ij},
\end{equation}
\begin{equation}
  \phi_{ij} = -\frac{2\pi}{\Phi_0} \int_{{\bf r}_i}^{{\bf r}_j} {\bf A}d{\bf r}
            = - \frac{2\pi B}{\Phi_0} \bar{x}_{ij} \; \Delta{y}_{ij},
\end{equation}
where the last expression holds for the Landau gauge 
for the vector potential ${\bf A}=(0,Bx)$.  

In this gauge 
the phase appears for the transfer involving a shift along $x$, which 
repeats itself with a translation of the unit cells along $x$ by 
$N_{\rm cell}$, where $N_{\rm cell}$ is the smallest $N$ that makes 
$N (q/p)(\Delta{y}_{ij}/L_y)$ 
an integer for all the bonds $\langle ij\rangle$ 
within or across a unit cell.
Thus we can perform a band-structure calculation 
regarding the $(N_{\rm cell} L_x, L_y)$ system as a new unit cell.  
Its size depends by construction not only on $q/p$ but also on the 
atomic configuration in the original unit cell of the 
superstructure (via $\Delta{y}_{ij}$).  
In this respect the magnetic cell defined here 
differs from those appearing in the magnetic translation group.  
The existence of the cell implies that the Brillouin zone will be 
$N_{\rm cell}$-folded.

Figure 1 displays 
the Hofstadter butterfly for simple realizations of 
Lieb's, 
Mielke's, and 
Tasaki's models, all assumed to have the 
square symmetry for simplicity. 
We can immediately see that, apart from the splitting 
of each dispersive band into magnetic minibands, 
the flat band in Lieb's case remains flat, 
whereas the interference-originated flat bands develop into 
peculiar butterflies as $B$ is increased.  

The fact that the topological flat bands 
can evade the effect of $B$ is analytically shown.  
There we have only to solve 
three simultaneous 
eigenequations for 
three amplitudes 
$\psi_A, \psi_B, \psi_C$ (A at the vertex and B,C at the mid points 
in the unit cell, 
depicted in Fig.1(a)). 
If we eliminate $\psi_B$ and $\psi_C$ the equation for nontrivial 
solutions for $\psi_A$ reduces to the 
corresponding equation for a simple square lattice 
if we translate $E_{\rm square}$ into $E^2-4$.  
On top of these there is a class of $E=0$ states that have 
$\psi_A\equiv 0$, so that we have indeed a flat band with its 
energy pinned at the original energy that is 
literally sandwiched by two butterflies 
mapped via $\pm (E_{\rm square}+4)^{1/2}$.  
Here the atomic level is taken to 
be $E=0$ (which coincides with $E_F$ when half-filled, i.e., one 
electron per atom) with $t=-1$.  

For $B=0$ the most compact `Wannier state' (that 
cannot be orthogonalized as stressed) on the flat band 
is as depicted in inset of Fig.1(a).  
In quantum chemical language these states correspond, for finite 
molecules, to `non-bonding 
molecular orbitals'.  
We can extend this by inspection to $B\neq 0$ as displayed in Fig.2.  
Curiously, the application of $B$ acts to deform the $E=0$ states into 
`elongated ring states' along $x$ (or $y$) 
in the Landau gauge, whose 
length equals to $p$ for $\tilde{B}=q/p$.  
This sharply contrasts with the usual Bloch-Landau state 
having the size of the magnetic length $\propto 1/\sqrt{B}$. 

On the other hand, it is not surprising the flatness is 
lost even for an infinitesimal $B$ 
in Mielke's or Tasaki's models, which 
rely heavily on a 
exact tuning of the interference.  
If we look more closely at their Hofstadter butterflies, 
we can show some symmetries 
such as (i) a full periodicity is accomplished when 
the magnetic flux 
penetrating the smallest loop in 
the lattice becomes $\Phi_0$ (which is reminiscent 
of the AB effect in the mesoscopic conductance\cite{webb}, 
(ii) there is a two-fold symmetry about 
$\tilde{B}=1$ for Mielke's model or a mirror symmetry 
about $\tilde{B}=2$ for Tasaki's.  

A more essential question is: can we identify the 
butterfly developed from the flat band with usual 
Landau's quantization?  
An indication that the situation is anomalous 
can be seen in Fig. 3, which 
displays how the `Landau quantization' looks like 
for Tasaki's model.  
For $\tilde{B}=q/p$ with $p$ even, we have a series of Landau's bands 
(Harper-broadened Landau levels in a nonparabolic bands) 
that have a zero gap at 
the position, $E_0=-2$, of the original flat-band. 
For an odd $p$ the gap vanishes.  

The density of states at $E_0$ thus alternates between zero and 
finite according to the parity of $p$, so that the 
orbital magnetic moment, $M=-\partial E_T /\partial B$ 
with $E_T$ being the total energy, becomes ill-defined 
along with the magnetic susceptibility.  
Another observation is that the density of states around $E_0$ 
spreads both below and above $E_0$ when $B$ is 
turned on, so that the total energy {\it decreases} when the 
magnetic flux is introduced if we start from a 
flat band less than half-filled.  
This might lead to an orbital ferromagnetism (a spontaneous induction of 
a network of `persistent currents'), 
although it has been pointed out\cite{alexandrov} 
in the context of the flux phase\cite{flux} in correlated 
electrons 
that a more accurate estimate of energy has to include 
the diamagnetic shift and the shrinkage of atomic orbitals.

We now turn to superhoneycomb systems, where in the classification 
by Shima and Aoki 
a class B$_0$ (B$_C$) system 
has to have, when bipartite, at least three (one) flat band(s) 
in the gap of semiconducting (semimetallic) bands.  
To define the classes, we can first 
note that a unit cell in a honeycomb system may be regarded as 
comprising two atomic clusters (or `superatoms'), 
where the two do not (case A) or have to (B) share an atom.  
The center of each superatom (a three-fold axis) may (case C) 
or may not (0) coincide with the position of an atom.  

The result for the Hofstadter butterfly (Fig.4) shows that the flat bands 
remain flat for $B>0$ 
no matter whether they are non-degenerate (B$_C$) or 
degenerate (B$_0$) at $B=0$.  
For the system depicted in Fig.4(a) 
the flat band is sandwiched between the 
butterfly for the simple honeycomb lattice\cite{rammal} 
just as in Fig.1(a), where the only difference is that 
the butterflies are now mapped via $\pm (E_{\rm honeycomb}+3)^{1/2}$.


The presumed superstructures are surprisingly stable against the 
band Jahn-Teller type distortion as seen from 
the total energy calculation\cite{shimaf}. 
For actual fabrication, one possibility would be 
to polymerize self-aligned 
organic molecules as realized in the van der Waals epitaxy\cite{koma}. 
Magneto-transport in these systems will be also of interest 
as in 3D organic materials\cite{iye}

We thank Koichi Kusakabe, Kazuhiko Kuroki, and Naoto Nagaosa 
for valuable discussions.


\newpage
\begin{center}
Figure captions
\end{center}

\vspace*{1cm}

{\bf Fig.1} Hofstadter's diagram 
(energy spectrum against $\tilde{B}$) 
for Lieb's(a), Mielke's(b), and 
Tasaki's(c) models, whose lattice structures are attached 
with $t$ etc being the transfer.  
Arrows here and figures below represent 
the position of the flat bands for $B=0$, 
while a `Wannier state' is encircled for Lieb's model. 
The spectrum are shown here for 
$\tilde{B}\equiv q/p$ with typically $p\leq 30$ or $1\leq q \leq 119$ with $p=120$.
\par
\ \\

{\bf Fig.2} 
An example of the $E=0$ 
`elongated ring states' in the Landau gauge, whose 
length equals to $p$ for $\tilde{B}=q/p (=1/5$ here).  
A circle represents a finite amplitude, while 
arrows indicate the phase.  
\par
\ \\

{\bf Fig.3} 
Typical band structures (projected onto $k_x$ plane or $k_y$ 
plane for finite numbers of ${\bf k}$'s) are displayed for $\tilde{B}=q/p$ 
with $p$ even ($q/p=1/10$, a) 
or $p$ odd ($q/p=1/5$, b) 
in Tasaki's model.  
Note a change in the vertical scale between (a) and (b).\par
\ \\

{\bf Fig.4} 
Hofstadter's diagram for class B$_0$ (a) 
or B$_C$ (b) superhoneycomb systems. \par

\end{document}